\documentclass[twocolumn]{article}
\usepackage{graphics}

\begin{document}

\title{On the nature of the superconducting gap in the cuprates}

\author{J. Quintanilla\(^{*,\dagger}\) and B.L. Gy\"orffy\\
University of Bristol, H.H. Wills Physics Laboratory,\\
Tyndall Av., Bristol BS8 1TL, U.K.}

\date{}

\maketitle

\begin{abstract} 
Recent experiments indicate that the excitation spectrum of the cuprates
is characterised, in the superconducting state, by two energy scales: the
{}``coherence energy'' \( \Delta _{c} \) and the {}``pseudogap'' \( \Delta
_{p} \). Here we consider a simple generalisation of the BCS model that
yields exotic pairing and can describe, phenomenologically, the generic
trends in the critical temperature \( T_{c} \) of cuprate superconductors.
We use the model to predict the gap in the single-particle spectrum
arising from the superconductivity and we find evidence that it
corresponds to the lower of the two energy scales, \( \Delta _{c} \), seen
in the experiments. This provides further support to the view that the
origin of the pseudogap is not superconducting fluctuations.
\end{abstract}

\section*{Introduction}

In this paper we comment on the nature of the superconducting gap
in the cuprates. Recently, Deutscher has identified two distinct energy
scales at low temperatures (in the superconducting state), which he
calls the coherence energy \( \Delta _{c} \) and the single-particle
excitation energy \( \Delta _{p} \) \cite{Deutscher-99}. The former
corresponds to the superconducting gap seen in phase-sensitive experiments
(Andreev reflection, penetration depth and Raman scattering), while
the later is the {}``pseudogap'' seen in experiments that probe
the one-particle spectrum (photoemission and tunnelling). In fact
some experiments belonging to the later class seem to be able to detect
both {}``gaps'' simultaneously: specifically, \( \Delta _{c} \)
and \( \Delta _{p} \) appear to have been resolved, as two distinct features
in the single-particle excitation spectrum, in angle-integrated
photoemission experiments on LSCO by Sato et al.~\cite{Sato-Yokoya-Naitoh-Takahashi-Yamada-Endoh-99}
and in intrinsic c-axis tunneling spectrosopy experiments on BSCCO
by Krasnov et al.~\cite{Krasnov-Yurgens-Winkler-Delsing-Claeson-00}
Here we address the following question: {}``which, if any, of these
two energies is the `superconducting gap', in the BCS sense?''

Since the issue that we wish to explore is fairly general, we use a
minimal model which might be expected to contain the relevant
physics. The basic idea is to exploit the fact that certain
properties of cuprate superconductors seem to be approximately
material-independent. In particular, the critical temperature \(
T_{c} \) \cite{Tallon-et-al-95} as well as the two energy scales \(
\Delta _{c} \) and \( \Delta _{p} \) \cite{Deutscher-99}, when
normalised to the critical temperature at optimal doping, \(
T_{c,max} \), appear to be fairly universal functions of the number
of holes per \( \textrm{CuO}_{2} \) unit, \( n_{h} \). Thus one
could expect that a very simple model which captures the essential
physics in these systems can describe this behaviour. More realistic
models would only be required to calculate material-specific
properties, such as \( T_{c} \) (as opposed to \( T_{c}/T_{c,max}
\)) accurately. In a sense the rationale behind our approach is the
same as in the BCS model which, with a very simplified description
of the electronic structure and the electron-electron interaction,
can nevertheless predict the isotope effect and the constant value
of the {}``BCS ratio'' \( \Delta _{0}/k_{B}T_{c} \) very
accurately, for most of the conventional superconductors. 
Unfortunately our model of the electron-electron interaction in the
cuprates is only pehnomenological, and hence we shall have little
to say about the nature of the  physical mechanism giving rise to
pairing. Nevertheless, as we  shall show presently the model has a
number of physical features  which if confirmed, by comparison with
experiments, will  shed some light on the central problem of the
high-\( T_{c} \) phenomena. In particular, the model that we shall
use has been showed to describe quantitatively the rise and fall of
\( T_{c}/T_{c,max} \) \cite{Quintanilla-Gyorffy-00}. Moreover, it
facilitates the investigations of superconductivity in both the BCS
and Bose-Einstein (BE)
limits \cite{Quintanilla-Gyorffy-Annett-Wallington-01}. The central
new result, presented here, is that the prediction of the model for
the superconducting gap in the single-particle excitation spectrum,
\( \Delta_{0} \), coincides, quite accurately, with the
\emph{lower} of the two energy scales identified by Deutscher,
namely the ``coherence energy'' \( \Delta_{c} \). On the basis of
this we argue that, although the model is too simple to describe
the superconducting state of the cuprates in detail, some of its
properties may generically characterise these materials. In short,
it provides evidence that \( \Delta _{c} \) can be understood as a
BCS-like \( d_{x^{2}-y^{2}} \)-wave gap, while the other energy
scale, \( \Delta _{p} \), seems to be quite unrelated to
superconductivity. Interestingly similar conclusions were reached
in ref.~\cite{Krasnov-Kovalev-Yurgens-Winkler-00} on the basis of
experimental investigations. Moreover, our approach, based on the
universal properties of cuprate superconductors, complements the
one recently adopted by Szotek, BLG and
Temmerman \cite{Szotek-Gyorffy-Temmerman-01}, who used a detailed
8-band model of the electronic structure of a specific material
(YBCO) to study the same question, reaching similar conclusions.

\section*{The model}

Our model consists of free electrons (holes) of effective mass \( m^{*} \)
interacting \textit{via} the attractive {}``delta-shell'' potential
(between opposite spins)

\begin{equation}
\label{delta-shell interaction}
V\left( \left| \mathbf{r}\right| \right) =-g\delta \left( \left| \mathbf{r}\right| -r_{0}\right) 
\end{equation}
This is, arguably, the simplest generalisation of the BCS model that
can describe exotic pairing \cite{Quintanilla-Gyorffy-00,Quintanilla-Gyorffy-Annett-Wallington-01}. 
The four parameters of the delta-shell model are the {}``coupling constant''
\( g \) (which has dimensions of \( energy\times length \)), the
distance at which the attraction takes place \( r_{0} \), the effective mass \(
m^{*} \), and the
density of charge carriers, per unit volume, \( n \).

Although formally similar to the {}``contact potential'' of BCS
theory, the physics described by the delta-shell potential are very
different: it corresponds to non-retarded attraction at a finite distance
\( r_{0} \). In that sense, it is more similar to the Hubbard-like
models with nearest-neighbour attraction \cite{Micnas-Ranninger-Robaskiewicz-90}.
But note that in the delta-shell model the distance \( r_{0} \) at
which the electrons attract each other is a free parameter that can
be varied continuously, and the non-interacting dispersion relation
is that of free electrons with an effective mass \( m^{*} \). 

A preliminary analysis of this model has been carried out using a
standard functional integral technique. The details of these calculations
are to be found in  ref.~\cite{Quintanilla-Gyorffy-Annett-Wallington-01},
however some of the results will be summarised in what follows as the
need arises and, for conveninence, the relevant formulae have been reproduced
in the appendix.

\section*{On the microscopic origin of the universality of the \protect\( T_{c}\protect \)
vs doping law}

\begin{figure}
{\centering \resizebox*{1\columnwidth}{!}{\includegraphics{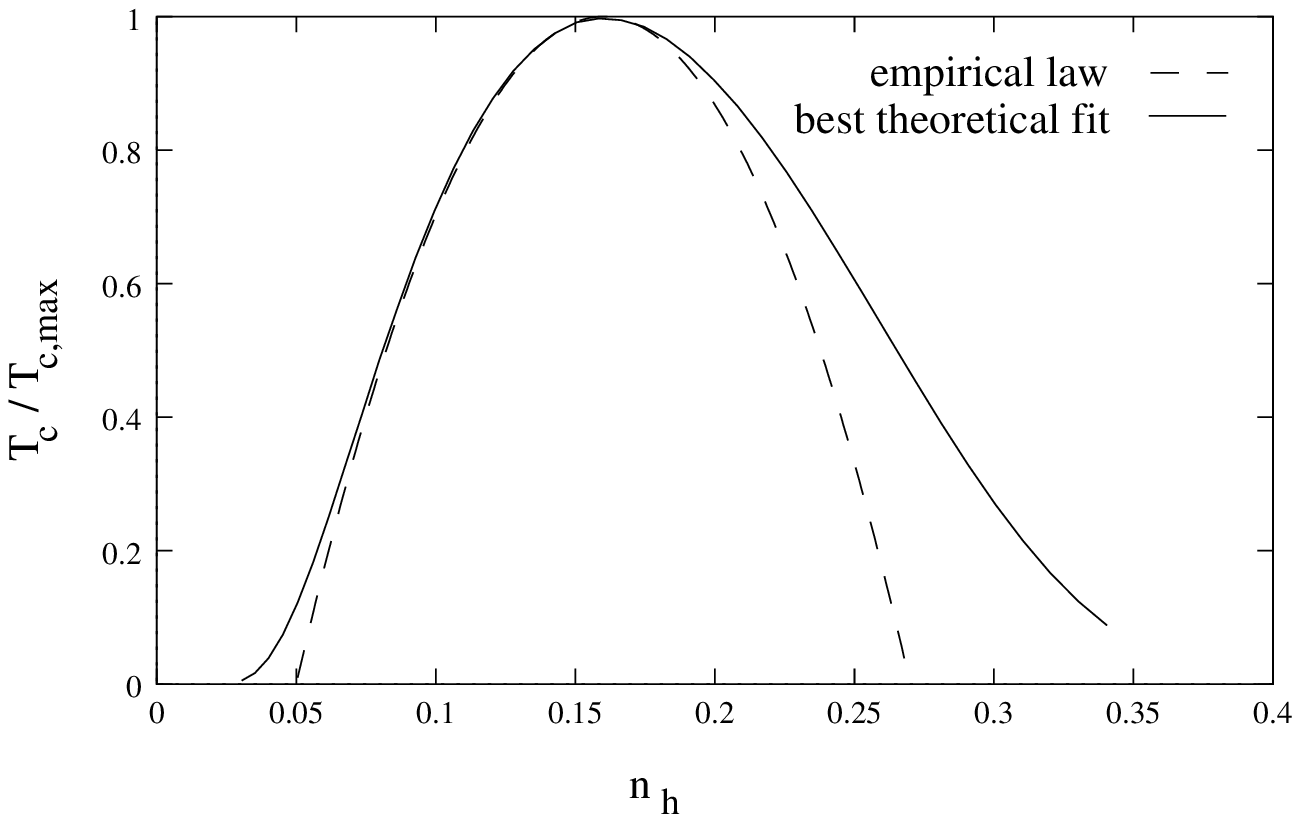}} \par}
\caption{The best fit of the rise and fall of the \protect\( d\protect \)-wave
value of \protect\( T_{c}/T_{c,max}\protect \) to the empirical law.\label{quigyof1}}
\end{figure}

A remarkable feature of the experimental data on the superconducting
cuprates is the dependence of \( T_{c}/T_{c,max} \) on the number
of holes, \( n_{h} \), per CuO\( _{2} \) \mbox{unit}
\cite{Tallon-et-al-95,Schon-et-al-01}. Namely, \( T_{c}/T_{c,max} \)
rises and falls as \( n_{h} \) increases from below optimal doping,
\( n_{h,max} \), to above, and the functional form of this relation
is the same for the different materials \cite{Tallon-et-al-95}. Explicitely,
it is given by \begin{equation}
\label{eq5}
T_{c}/T_{c,max}=\left[ 1-82.6(n_{h}-0.16)^{2}\right] 
\end{equation}
Remarkably, in the delta-shell model, the ratio \( T_{c}/T_{c,max} \) does not
depend on \( g \), \( r_{0} \), \( m^{*} \) and \( n \) independently, but only
on the two dimensionless quantities \( \tilde{n} \equiv n \frac{4\pi}{3}
r_{0}^{3} \) and \( \tilde{g}\equiv \frac{g}{r_{0}}/ \frac{\hbar
^{2}}{2m^{*}r_{0}^{2}} \).\label{def-g-n} The dimensionless density \( \tilde{n} \) can be
related to \( n_{h} \) through the value of the ratio \( r_{0}/r_{\rm {Cu-Cu}}
\), where the effective Cu--Cu distance,  \( r_{\rm{Cu-Cu}} \), is defined in
terms of the proportionality between  the number of holes per \mbox{CuO\( _{2}
\)} unit and their density per unit volume: \( n_{h}=n r_{\rm {Cu-Cu}}^{3}
\). Thus the values of the dimensionless coupling constant \( \tilde{g} \)
and the rescaled range of the attraction, \( r_{0}/r_{\rm {Cu-Cu}} \),
completely determine, for our model, the dependence of \( T_{c}/T_{c,max} \) on
\( n_{h} \). We find that, as in the cuprates, \( T_{c}/T_{c,max} \) rises and
falls as a function of \( n_{h} \), for fixed \( \tilde{g} \) and \( r_{0}/r_{\rm{Cu-Cu}} \). Moreover, as we have pointed out
earlier \cite{Quintanilla-Gyorffy-00}, we can choose these two parameters so that
this rise--and--fall reproduces the shape of the experimental curve, an
inverted parabola defined by the position of its maximum and its width. The
condition that \( \tilde{g} \) and \( r_{0}/r_{\rm{Cu-Cu}} \) are fixed to the
values that produce the best fit, in the \( d \)-channel (Cooper pairs with
angular momentum quantum number \( l=2 \)), is equivalent to the following two
relations between the four parameters \( g \), \( r_{0} \), \( m^{*} \) and \(
r_{\rm{Cu-Cu}} \):
\begin{equation}
\label{parameter-values-1}
r_{0}=2.2 \,\, r_{\rm {Cu-Cu}}
\end{equation}
 and \begin{equation}
\label{parameter-values-2}
g/r_{0}=0.6 \,\, \hbar ^{2}/2m^{*}r^{2}_{\rm {Cu-Cu}}
\end{equation}
Note that \( m^{*} \) and \( r_{\rm{Cu-Cu}} \) are normal state properties 
independent of the attractive interaction in eq. (\ref{delta-shell interaction}). 
For clarity, the corresponding fit is recorded
in figure \ref{quigyof1}. Evidently, the above relations
determine \( g \) and \( r_{0} \) for each system:
YBCO, LSCO, etc.,  characterised by the frankly phenomenological, but
neverteless clearly  material-specific, parameters \( m^{*} \) and \(
r_{\rm{Cu-Cu}} \). Thus although the four parameters are independent
of doping, their fixed values are different for different systems. The
universal character of the empirical law of eq. (\ref{eq5}), on the other hand,
is embodied in the fact that the relations
(\ref{parameter-values-1},\ref{parameter-values-2}) between the four quantities
are the same for all systems. By  fitting relative quantities like \(
T_{c}/T_{c,max} \) and universal  trends with doping we can hope to have
obtained results dependent more on  the overall properties of our model than on
its material-specific, no doubt inadequate, details. 

We would also like to stress that other, \( l\neq 2 \), singlet pairing
channels could not fit the oberved dependence of \( T_{c}/T_{c,max} \)
on the doping level. In particular, for pairs with angular momentum
quantum number \( l=0 \) (\( s \)-wave), the critical temperature
is finite for arbitrarily small values of the density, in contradiction
with the observed behaviour, while other pairing channels are negligible
in the range of densities of interest \cite{Quintanilla-Gyorffy-Annett-Wallington-01}.
Thus the interpretation of the observed rise-and-fall of \( T_{c} \)
with doping in terms of the delta-shell model implies \( d \)-wave
symmetry in agreement with the current consensus \cite{Annett-Goldenfeld-Leggett-96}.

\section*{Pressure dependence of \protect\( T_{c}\protect \)}

As a further validation of our model we will consider, briefly, the
dependence of \( T_{c} \) on pressure. 
Experimentally, there are two contributions to the variation
of \( T_{c} \) with applied hydrostatic pressure \cite{Wijngaarden}. The extrinsic contribution
is caused by a pressure-induced change of the doping
level, combined with the universal \( T_{c} \) vs doping law (\ref{eq5}). On
the other hand, there is also  an \emph{intrinsic} enhancement of \(
T_{c} \), i.e. an increase of the value of \( T_{c,max} \).
For the values of the parameters
given in eqs. (\ref{parameter-values-1},\ref{parameter-values-2}),
\( T_{c,max} \) scales with \( \hbar ^{2}/2m^{*}r_{\rm {Cu-Cu}}^{2} \)
and is given by \begin{equation}
\label{Tcmax}
k_{B}T_{c,max}\approx 0.1\, \, \hbar ^{2}/2m^{*}r_{\rm {Cu-Cu}}^{2}
\end{equation}
 Thus although \( T_{c}/T_{c,max} \) is universal, in the sense that
it only depends on \( n_{h} \), the absolute value of \( T_{c} \)
is material-specific, as one would expect in view of the experimental
data \cite{Ginsberg-Phys-Prop}, because it depends also on \( m^{*} \)
and \( r_{\rm {Cu-Cu}} \). Although, along the lines set in the introduction,
we do not expect this simple model to reproduce the absolute values
of \( T_{c,max} \) accurately, if it contains any of the relevant
physics \mbox{eq. (\ref{Tcmax})} must be compatible with the values of \( T_{c,max} \)
observed experimentally. Notably, in the cuprates, \( T_{c,max} \)
can go from \( \sim 30\rm {K} \) to \( \sim 150\rm {K} \), which
is consistent with our prediction if \( r_{\rm {Cu-Cu}}\sim 5{\rm\AA} \)
and the effective masses are of the order of a few times the bare
electron mass. Moreover eq. (\ref{Tcmax}) provides a simple interpretation
for the origin of the \emph{intrinsic} enhancement of the critical temperature: 
it is due to the increase of the {}``localisation
energy'' \( \hbar ^{2}/2m^{*}r_{\rm {Cu-Cu}}^{2} \) associated with
the reduction in size of \( r_{\rm {Cu-Cu}} \). 
By taking the derivative
of \( T_{c,max} \) with respect to the volume of the unit cell \( abc \)
(which is proportional to \( r_{\rm {Cu-Cu}}^{3} \)) we can obtain
an estimate of such enhancement: \begin{equation}
\label{eq9}
\left. \frac{\partial \ln T_{c,max}}{\partial P}\right| _{P=0}\approx \left[ \frac{2}{3}+\frac{\partial {\ln m^{*}}}{\partial \ln (abc)}\right] k_{V}
\end{equation}
 Here \( k_{V} \equiv -\partial \ln (abc) / \partial P\) is the compressibility of the material. Note that
in writing (\ref{eq9}) we have assumed that the relations (\ref{parameter-values-1},\ref{parameter-values-2})
do not depend on applied pressure. This would be consistent with the
universality of (\ref{parameter-values-1},\ref{parameter-values-2})
if that universality were exact, but note that such universality is
probably only approximate. Even so, we can easily use (\ref{eq9})
to obtain a specific value as follows. We can employ the usual way
of estimating the volume dependance of the electronic effective mass,
which is of interest in many fields of solid state physics, by an
appeal to the relationship between \( \hbar ^{2}/2m^{*} \) and an
appropriate hopping integral, \( t \), in the tight-binding description
of the relevant band. 
In the simplest case of 
a cubic lattice with constant \(r_{\rm Cu-Cu}\) this is 
\(m^* \sim 1/t r_{\rm Cu-Cu}^2\). 
Then we use the dependance of \( t \) on the
nearest-neighbour distance to estimate the changes in \( m^{*} \)
with changes in the lattice parameters. Thus we may use 
\( t\sim r_{\rm {Cu-Cu}}^{-\gamma } \)
and, assuming that we are dealing with \( d \)-orbitals, take \( \gamma =5 \) \cite{Harrison-80}.
This yields \begin{equation}
\label{eq10}
\frac{\partial \ln T_{c,max}}{\partial P}\approx \frac{5}{3}k_{V}
\end{equation}
 Unfortunately, we are not aware of many measurements of the variation
of the \emph{maximum} value of the critical temperature under applied
hydrostatic pressure. However Wijngaarden \textit{\emph{}}\textit{et
al.} \cite{Wijngaarden} have studied the single layer compound \( {\textrm{Tl}_{0.5}\textrm{Pb}_{0.5}\textrm{Sr}_{2}\textrm{Ca}_{1-x}\textrm{Y}_{x}\textrm{Cu}_{2}\textrm{O}_{7}} \)
and from their data we deduce \( \left( \partial \ln T_{c,max}/\partial P\right) _{P=0}\approx 12.3\, \, 10^{-3}\, \, {\textrm{GPa}}^{-1} \).
Using the value of the compressibility calculated by Cornelius \textit{et
al.} \cite{Cornelius}, \( k_{V}=8.4\, \, 10^{-3}\, \, {\textrm{GPa}}^{-1} \),
and our formula in equation (\ref{eq10}), we predict \( \left( \partial \ln T_{c,max}/\partial P\right) _{P=0}\approx 14.0\, \, 10^{-3}\, \, {\textrm{GPa}}^{-1} \),
in reasonable agreement with the experiment. Similarly, Schlachter
\textit{et al.} \cite{Schlachter-et-al-99} have obtained \( \left( \partial T_{c,max}/\partial P\right) _{P=0}\approx 0.8\, \, {\textrm{K}}{\textrm{GPa}}^{-1} \)
for \( {\textrm{YBa}_{2}\textrm{Cu}_{3}\textrm{O}_{x}} \) which,
using \( T_{c,max}\approx 92{\textrm{K}} \), yields \( \left( \partial \ln T_{c,max}/\partial P\right) _{P=0}\approx 8.7\, \, 10^{-3}\, \, {\textrm{GPa}}^{-1} \).
The theoretical prediction in this case, using the value \( k_{V}=8.1\, \, 10^{-3}\, \, {\textrm{GPa}}^{-1} \)
quoted in ref.~\cite{Cornelius}, is \( \left( \partial \ln T_{c,max}/\partial P\right) _{P=0}\approx 13.5\, \, 10^{-3}\, \, {\textrm{GPa}}^{-1} \).
Thus the \emph{intrinsic} enhancement of the critical temperature
with applied hydrostatic pressure
can be understood essentially, at the qualitative level, as the result
of the change of the length \( r_{\rm {Cu-Cu}} \), and therefore
of the corresponding localisation energy \( \hbar ^{2}/2m^{*}r_{\rm {Cu-Cu}}^{2} \),
under pressure.\footnote{%
Note that in deriving this result we have used that \( k_{B}T_{c,max}\propto \hbar ^{2}/2m^{*}r_{\rm{Cu-Cu}}^{2} \),
which is more general than (\ref{Tcmax}) and, in fact, would hold
for any model for which \( \hbar^{2}/2m^{*}r_{\rm{Cu-Cu}}^{2} \)
is the relevant energy setting the scale of \( T_{c,max} \).
}

We end by stressing that the results presented in this section
(unlike those in the previous and following ones) refer to
material-specific properties of the cuprates and therefore, in
accordance with the philosophy outlined in the Introduction, one
must not expect our simple model to provide a detailed description
of them. For example, it can not describe the observed effect of the
number of adjacent CuO\( _{2} \) planes per unit cell, nor of
applied uniaxial stress \cite{Locquet-et-al-98}, on \( T_{c} \),
since obviously the parameters \( r_{{\rm Cu-Cu}} \) and \( m^{*} \)
do not describe the varying degrees of anisotropy of the normal
states.

\section*{The superconducting gap}

We shall now return to the main topic of this paper. On the basis
of the discussion in the previous sections, we can expect that the
delta-shell model, together with the relations (\ref{parameter-values-1},\ref{parameter-values-2}),
can describe quantitatively other material-independent properties
of the superconducting state of the cuprates. In fact
%, since the relevant parameters have been fixed, 
it \emph{predicts},
with no adjustable parameters, the value of any such quantities. The
most interesting one is the ratio between the energy gap to one-particle
excitations, at zero temperature, and the critical temperature at
optimal doping: \( 2\Delta _{0}/k_{B}T_{c,max} \). In the model,
this quantity is universal, in the same sense as \( T_{c}/T_{c,max} \),
because \( \Delta _{0} \) also scales with \( \hbar ^{2}/2m^{*}r_{\rm {Cu-Cu}}^{2} \). 

In a very general context, there are two distinct scenarios, which we
shall now briefly recall (see \cite{Randeria-95}, and references within).
In the BCS regime (\( \mu >0 \)) the fermionic quasiparticles with
minimum excitation energy lie on a surface in \textbf{\( \mathbf{k}
\)}-space that is defined by \( \hbar
^{2}\mathbf{k}_{min}^{2}/2m^{*}=\mu  \) and the corresponding gap is \(
2\Delta _{0}=2\left| \Delta \left( \mathbf{k}_{min}\right) \right|  \),
where \( \Delta \left( \mathbf{k}\right)  \) is the usual BCS {}``gap
function''. The extreme case of this is the BCS \emph{limit}, defined by
\( \mu \gg \left| \Delta \left( \mathbf{k}\right) \right|  \) for all
\textbf{\( \mathbf{k} \)}, in which the chemical potential is
approximately equal to the Fermi energy, \( \mu \approx \varepsilon _{F}
\), and so the fermionic quasiparticles with minimum energy have wave
vector \( \mathbf{k}_{min}\approx \mathbf{k}_{F} \), i.e. they lie in the
Fermi surface. In the opposite, BE regime (\( \mu <0 \)), quasiparticles
with minimum excitation energy have \( \mathbf{k}=0 \) and the energy gap
to single-particle (fermionic) excitations is \( 2\Delta _{0}=2\sqrt{\mu
^{2}+\left| \Delta \left( 0\right) \right| ^{2}} \), which, in the BE
\emph{limit} (\( \mu \ll -\left| \Delta \left( \mathbf{k}\right) \right| 
\)), becomes the binding energy \( \varepsilon _{b} \) of a pair: \(
2\Delta _{0}\approx 2\mu \approx \varepsilon _{b} \). However in the BE
regime the lowest-lying excitations are not fermionic, but bosonic, and
therefore there is a second, smaller energy scale, which is associated
with the break-up of coherence in the condensate without the dissociation
of any pairs, whose value could be related to \( \Delta \left(
\mathbf{k}\right)  \). 

The delta-shell model can display both BCS-like and BE-like
behaviour, depending on the values of the dimensionless parameters
\( \tilde{g} \) and \( \tilde{n} \) introduced above. Specifically,
for the values that correspond to eqs.
(\ref{parameter-values-1},\ref{parameter-values-2}), the model is
close to the BCS limit
\cite{Quintanilla-Gyorffy-Annett-Wallington-01}. In fact the
rise-and-fall of \( T_{c} \) with the doping level that we have
found in the delta-shell model is only present at low values of \(
\tilde{g} \). At higher values of the dimensionless coupling
constant \( \tilde{g} \), the critical temperature becomes a
monotonic function of the density, which furthermore has a finite
value for any finite value of \( n \), as it tends towards the
Bose-Einstein condensation temperature \( T_{c}^{BE}\propto n^{2/3}
\). Moreover the rotational symmetry of the model is restored in
this limit due to the fragmentation of the condensate. Thus on the
basis of the discussion presented in the previous section we
conclude the first of the above scenarios to be operating in the
cuprates. Namely, we expect there to be a single energy scale \(
\Delta_{c} \) coming from the superconductivity, which would
manifest itself both as the characteristic energy seen in phase
sensitive experiments and as a gap in the single-particle excitation
spectrum. In this picture, the origin of the larger {}``pseudogap''
energy \( \Delta_{p} \) would be quite independent from
superconductivity. 

\begin{figure}
{\centering \resizebox*{1\columnwidth}{!}{\includegraphics{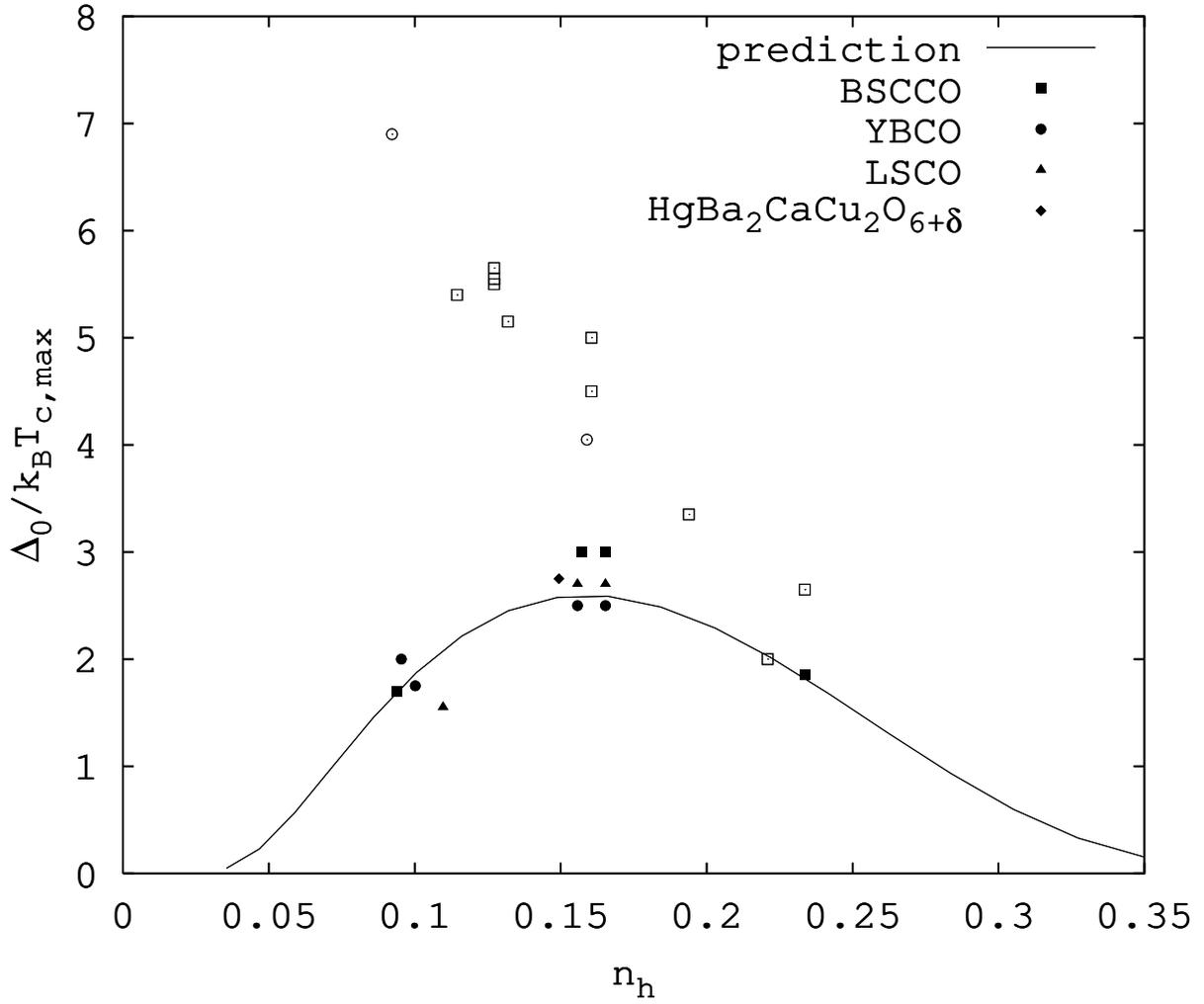}} \par}
\caption{The theoretical prediction for the superconducting gap in the single-particle
spectrum \protect\( \Delta _{0}\protect \) compared to the experimental
data for the {}``coherence energy'' \protect\( \Delta _{c}\protect \)
(filled symbols) and the {}``pseudogap'' energy \protect\( \Delta _{p}\protect \)
(open symbols) from ref.~\cite{Deutscher-99}. All quantities are
normalised to the critical temperature at optimal doping, \protect\( T_{c,max}\protect \).\label{gd}}
\end{figure}

Figure \ref{gd} shows the experimental data from ref.~\cite{Deutscher-99}
for the two energy scales \( \Delta _{c} \) and \( \Delta _{p} \),
together with our theoretical prediciton for the maximum value of
the superconducting energy gap in the single-particle spectrum across
the Fermi surface, \( \Delta _{0} \), for the purported \( d_{x^{2}-y^{2}} \)
symmetry of the order parameter \cite{Annett-Goldenfeld-Leggett-96}.
All quantities are normalised to \( k_{B}T_{c,max} \), which is the
way in which, both in the model and, approximately, in the experiments,
their values become material-independent. Remarkably, the agreement
between \( \Delta _{0} \) and \( \Delta _{c} \) is rather good.
In particular the ratio of \( \Delta _{0} \) to \( k_{B}T_{c,max} \)
is \( \sim 5 \) at optimal doping, both in the theory and the experiments.
Obviously, because of the simplicity of our model it is not easy to
decide what to make of the above very good agreement. Nevertheless,
at the minimum we conclude that it provides evidence in favour of
the BCS-like scenario. Moreover the model seems again to confirm the
current consensus on the symmetry of the order parameter: for pairs
with \( d_{3z^{2}-r^{2}} \) symmetry,\label{d3z2-r2} \( 2\Delta _{0}/k_{B}T_{c,max}\approx 6.4 \)
at optimal doping, which is too high to fit the experiments (note
that, unlike at \( T_{c} \), at \( T=0 \) the different \( d \)-wave
channels are non-degenerate). Presumably, it is the presence of the
crystal field that makes one of the \( d \)-wave symmetries preferred
over the others. Since this is absent from our model all we can do
is examine each of the possible symmetries independently and compare
them to the experiments.

Of course, our interpretation of the data presented in ref.~\cite{Deutscher-99}
differs from that of the author. In effect, it was assumed at the
time that there was only one energy gap in the single-particle excitation
spectrum, corresponding to the {}``pseudogap'' energy \( \Delta _{p} \).
This was legitimate back then because only experiments that probed
the phase coherence of the condensate (Andreev, Raman, penetration
depth) had been able to measure \( \Delta _{c} \). However, as we
mentioned in the introduction, it has recently been shown experimentally \cite{Sato-Yokoya-Naitoh-Takahashi-Yamada-Endoh-99,Krasnov-Yurgens-Winkler-Delsing-Claeson-00}
that there are actually two gaps in the spectrum of single-particle
excitations: one of them corresponds to \( \Delta _{p} \), while
the other, smaller one closes in at \( T=T_{c} \), very much like
a BCS gap. It is this second gap that we have obtained in our calculations,
and in our picture it should coincide with the coherence energy \( \Delta _{c} \).
As regards the other energy scale, \( \Delta _{p} \), we have no
sign of it in our theory and hence we are inclined to conclude that
its origin is not superconductivity. In particular,  it can not be
due to {}``preformed pair'' fluctuations of the superconducting
order parameter which, as we have discussed, are ruled out on
the basis of the empirically determined values of the parameters, 
eqs. (\ref{parameter-values-1},\ref{parameter-values-2}),
and their interepretation in terms of the calculations 
presented in ref.~\cite{Quintanilla-Gyorffy-Annett-Wallington-01}.

\section*{Conclusion}

We have addressed the question of whether the coherence energy range
\( \Delta _{c} \), identified by Deutscher, at low temperatures, in
cuprate superconductors, as distinct from the {}``pseudo-gap''
energy \( \Delta _{p} \), can be understood as a superconducting gap
in the sense of BCS theory. Capitalising on the fact that certain
properties of cuprate superconductors are approximately
material-independent, we have studied a simple model, obtained as
the simplest generalisation of the \mbox{BCS} model compatible with
exotic pairing. It features free electrons with a non-retarded,
effective interaction potential which is attractive at a
well-defined finite distance \( r_{0} \) and is zero for all other
separations. This {}``delta-shell'' model provides a mechanism for
the rise and fall of \( T_{c} \) with doping, as a consequence of
the interplay between the carrier-carrier distance or, in other
words, the Fermi wavelength \( \left| \mathbf{k}_{F}\right| ^{-1} \)
and the new microscopic length scale \( r_{0} \). Morover, it can
correlate this behaviour with the \( d \)-wave symmetry of the order
parameter and it provides an interpretation of its approximate
universality in terms of certain relations between the parameters of
the model. On this basis, the model can be used to predict, with no
fitting parameters, the corresponding universal behaviour of the gap
in the quasiparticle spectrum due to the superconductivity. For the
empirically determined values of the parameters the model is in the
BCS regime, as opposed to that of  Bose-Einstein condensation of
``preformed pairs'', which leads to the prediction that the gap to
single-particle excitations arising from the superconductivity is
equal to \( \Delta _{c} \). In fact the prediction for that gap
agrees quite satisfactorily with the behaviour of \( \Delta _{c} \),
suggesting a different origin for the pseudogap \( \Delta _{p} \). 

The above conclusion is compatible with general arguments based on
the available experimental data which indicate that the descripition
of the charge carriers in the Bose-Einstein limit is inappropriate
\cite{Chakraverty-Ranninger-Feinberg-98}. Although a model that
accomodates this fact while still describing the pseudogap in terms
of pairing fluctuations has been devised \cite{Ranninger-00}, it
has also been suggested
\cite{Fleck-Alexander-Lichtenstein-Andrzej-OleS-01,Moshchalkov-Vanacken-Trappeniers-01}
that the pseudogap may follow from the spectral properties of stripe
phases
\cite{Tranquada-Sternlieb-Axe-Nakamura-Uchida-95,Mook-Dogan-99}.
More generally, a pseudogap may be a generic feature of systems that
are intrinsically inhomogeneous due to the competition between
ordered states separated by a first-order phase transition
\cite{Moreo-Yunoki-Dagotto-99,Burgy-Mayr-MartinMayor-Moreo-Dagotto-01}.
Such picture becomes especially appealing in the light of recent
scanning tunnelling microscope studies of BSCCO
\cite{Pan-et-al-01,Lang-et-al-02}. It could be investigated in our model
by evaluating the density fluctuations induced by the attractive
interaction, in addition to the pairing fluctuations considered in
the present work.

Our analysis complements the one carried out by Szotek, BLG and Temmerman \cite{Szotek-Gyorffy-Temmerman-01},
who also modeled \( \Delta _{c} \) as a BCS superconducting gap,
reaching similar conclusions. The main difference between our approach
and theirs is that, instead of focusing on a specific material using
a detailed description of the electronic structure, we have used a
very simple model to capture those features that are approximately
material-independent. Specifically, our calculations are not based
on a fit to the value of \( T_{c,max} \) (which can not be predicted
accurately by such a simple model), but rather to the width of the
\( T_{c}/T_{c,max} \) vs \( n_{h} \) curve, and the position of
its maximum, which are determined by the values of the range of the interaction
\( r_{0} \) and its strength \( g/r_{0} \), rescaled to the effective Cu--Cu
distance \( r_{{\rm Cu-Cu}} \) and the corresponding ``localisation energy''
\( \hbar^2 / 2m^{*}r_{{\rm Cu-Cu}}^2\), respectively. Our result that the model
is in the BCS limit for the empirically--determined values of these parameters
speaks in support of the methodology employed in ref.~\cite{Szotek-Gyorffy-Temmerman-01}, where this was assumed \emph{a priori} by
using a BCS-like mean-field theory to calculate \( T_{c} \).

Finally we stress that the rise and fall of the critical temperature
\( T_{c} \) and the superconducting gap \( \Delta _{0} \), in the
BCS limit, is probably a generic behaviour for attractive potentials
with a minimum at some finite separation. Essentially, the gap function
depends, in this case, on the \emph{magnitude} of \( \mathbf{k} \)
as well as on its direction: \( \Delta(\mathbf{k})=\Delta(\left|
\mathbf{k}\right| ,\hat{\mathbf{k}}) \),
and thus its value on the Fermi surface \( \mathbf{k}=\mathbf{k}_{F} \)
depends strongly on the size of the Fermi surface, given by \( \left| \mathbf{k}_{F}\right|  \).
This situation is quite different from the BCS model, and its usual \( d
\)-wave generalisations \cite[for example]{Balian-64}, for which the
gap function is independent of \( \left| \mathbf{k} \right| \). Note that the present
scenario for the rise and fall of \( T_{c} \) is quite different
from previous proposals. In particular, in our picture this rise and
fall is not due to the competition of superconductivity with other
kinds of order or the suppression of \( T_{c} \) by fluctuations
away from optimal doping. Moreover, the value of \( T_{c} \) does
not necessarily correlate with the DOS on the Fermi surface.\footnote{%
In the context of conventional, BCS superconductivity, this has usually
been thought to be the primary mechanism by which \( T_{c} \) can
have a non-monotonic dependence on doping. Interestingly, a rise and
fall of the critical temperature with doping, which does not seem
to correlate to the DOS, has been observed in experiments on gate
voltage-doped C\( _{60} \) films \cite{Schon-Kloc-Batlogg-00}.
} It could be tested if an experimental method to measure directly the 
variation of the gap function \( \Delta(\mathbf{k}) \) with \( \mathbf{k} \) 
in the radial direction of increasing \( \left| \mathbf{k}\right|  \),
rather than along the Fermi surface, e.g. one that would give the
value of \( \left( \partial \Delta(\mathbf{k})/\partial \left| \mathbf{k}\right| \right)_{\left| \mathbf{k}\right| =\left| \mathbf{k}_{F}\right| } \),
were available.

\section*{Acknowledgements}

We wish to thank James Annett, Jon Wallington and Klaus Capelle for
useful discussions, and Phillip Howell for detailed, constructive
criticism of an earlier version of the manuscript. BLG wishes to
thank the Institute of Theoretical Physics at the University of
California at Santa Barbara for hospitality during the preparation
of an earlier version of the manuscript. JQ acknowledges financial
support from the TMR programme (EU), \textit{via} a Marie Curie
fellowship (contract No.~ERBFMBICT983194) and, during the final
stages of the preparation of the manuscript, from FAPESP (Brazil),
\textit{via} a post-doctoral fellowship (process No.~01/10461-8).

\(^*\)Present address:  Instituto de F\'{\i}sica de S\~ao Carlos
(Grupo de F\'{\i}sica Te\'orica), Universidade de S\~ao Paulo,
Caixa Postal 369, 13560-970 S\~ao Carlos, SP, Brazil.

\(^{\dagger}\)Electronic address: quintanilla@if.sc.usp.br.

\appendix

\section*{Appendix}

In ref.~\cite{Quintanilla-Gyorffy-Annett-Wallington-01} the ground state
of the delta-shell model was studied in the mean-field approximation in
which the gap function \(\Delta({\bf k})\) and the chemical potential
\(\mu\) are determined by two coupled equations \cite{DeGennes-66}. The
theory is particularly transparent in terms of their dimensionless
counterparts \(\tilde{\Delta}_{\tilde{\bf
k}}\equiv\left(\frac{\hbar^2}{2m^*r_0^2}\right)^{-1}\Delta({\bf k})\) and
\(\tilde{\mu} \equiv \left(\frac{\hbar^2}{2m^*r_0^2}\right)^{-1} \mu\)
(where \(\tilde{\bf k} \equiv {\bf k}r_0\) is the dimensionless  wave
vector). If \(\tilde{\Delta}_{\tilde{\bf k}}\) is expanded in spherical
harmonics \(Y_{lm}(\hat{\bf k})\), it is found to have the form  
	\[
\tilde{\Delta}_{\tilde{\bf k}} = \sum_{lm} \tilde{\Delta}_{lm}
j_{l}(|\tilde{\bf k}|) Y_{lm}(\hat{\bf k}) 
	\label{general delta-shell gap function} 
	\]
i.e. the radial part is given by the spherical Bessel
functions \(j_l(|\tilde{\bf k}|)\), and the two equations
are, for singlet pairing (\(\tilde{\Delta}_{lm}=0\) for
\(l\) odd),
  \begin{eqnarray}
\tilde{\Delta }_{lm} & = & 
\tilde{g}\sum _{l'm'}
	\left\{ 
\int \frac{d^{3}\tilde{\mathbf{k}}}{\left( 2\pi \right) ^{3}}
\frac{\tilde{\Lambda }_{lm,l'm'}( \tilde{\mathbf{k}}) }
{2\sqrt{(\tilde{\bf k}^2-\tilde{\mu})^2+\tilde{\Delta}_{\bf k}^2}}
	\right\} \tilde{\Delta }_{l'm'}
\label{rescaled-gap-eq}
\\
\tilde{n} & = & \int \frac{d^{3}\tilde{\mathbf{k}}}{6\pi ^{2}}\left(
1-\frac{\tilde{\varepsilon }_{\tilde{\mathbf{k}}}}{\sqrt{(\tilde{\bf k}^2-\tilde{\mu})^2+\tilde{\Delta}_{\bf k}^2}}\right) 
\label{rescaled-n-eq}
\end{eqnarray}
where \(\tilde{g}\) and \(\tilde{n}\) are the dimensionless coupling
constant and density defined in the text, and
	\begin{equation}
\tilde{\Lambda }_{lm,l'm'}( \tilde{\mathbf{k}})   \equiv  
\left( 4\pi \right) ^{2}
j_{l}(|\tilde{\mathbf{k}}|) j_{l'}(|\tilde{\mathbf{k}}|) 
Y^{*}_{lm}( \hat{\mathbf{k}}) Y_{l'm'}( \hat{\mathbf{k}}) 
\label{rescaled-Lambda}
	\end{equation}
All the ground-state properties that we discuss in the text were obtained
by solving the above two equations self-consistently, for ``trial gound
states'' in which specific constraints were imposed on the
\(\tilde{\Delta}_{lm}\). Note that these equations are valid
approximations not only in the BCS limit, encountered for large
\(\tilde{n}\) or when \(\tilde{g}\) is below the threshold to  bind an
isolated pair, but also in the opposite, Bose-Einstein condensation
limit \cite{Leggett-80,Nozieres-SchmittRink-85}.

To find the condensation temperature \(T_c\) we went a small step beyond
the mean-field theory, including quadratic fluctuations about it. In a
manner similar to that reviewed in ref.~\cite{Randeria-95} we found that
\(\tilde{\beta}_c=\frac{\hbar^2}{2m^*r_0^2}/k_BT_c\) is determined by 
\begin{eqnarray}
\frac{1}{\tilde{g}}&=&\frac{2}{\pi }\int _{0}^{\infty
}d| \tilde{\mathbf{k}}| | \tilde{\mathbf{k}}|
^{2}j_{l}\left( | \tilde{\mathbf{k}}| \right)
^{2}\frac{1-2f\left[ \tilde{\beta }_{c}\left(|\tilde{\bf k}|^2-\tilde{\mu}_{c}\right)\right] }
{2\left(|\tilde{\bf k}|^2-\tilde{\mu}_{c}\right)}
\label{rescaled-tc-eq}
\\
\tilde{n}&=&
\frac{4}{3\pi }\int _{0}^{\infty }d|\tilde{\mathbf{k}}| 
| \tilde{\mathbf{k}}| ^{2}
\left\lbrace
f\left[ \tilde{\beta }_{c}\left(|\tilde{\bf
k}|^2-\tilde{\mu}_{c}\right)\right]
+g\left( \tilde{\beta }_{c}\frac{|\tilde{\mathbf{k}}|^{2}}{2}\right) 
\tilde{w}^{\tilde{\beta }_{c},\tilde{\mu }_{c}}_{l,| \tilde{\mathbf{k}}|} \right\rbrace
\label{density-eq-rescaled}
\end{eqnarray}
where \(\tilde{\mu}_c\) is the value of \(\tilde{\mu}\) at the transition
and \(f(z)\) and \(g(z)\) denote the usual Fermi and Bose distribution
functions. The term containing the Bose function, times the ``weight'' \(
\tilde{w}^{\tilde{\beta }_{c},\tilde{\mu }_{c}}_{| \tilde{\mathbf{k}}|}
\), is the contribution from the above mentioned fluctuations, described
here to the lowest non-trivial order in their frequency and momentum, and
it signals the presence of preformed pairs, with the same internal
angular momentum quantum number \(l\) as the condensate, above \(T_c\)
(in these expressions, the contribution from other pairs has been neglected). This contribution
dominates in the Bose-Einstein limit (where its ``weight'' takes the form
\( \tilde{w}^{\tilde{\beta }_{c},\tilde{\mu }_{c}}_{l,|
\tilde{\mathbf{k}}|} \approx 2l+1 \), and thus it simply accounts for the
degeneracy of the internal state of the pairs) while being negligible in
the BCS limit. Note that the critical temperature is degenerate in the
magnetic quantum number \(m\), although it still depends on  \(l\).

From the point of view of the present paper it is particularly relevant
that both \(\Delta_{lm}/k_BT_{c,max}\) and \(T_c/T_{c,max}\) depend only
on the dimensionless parameters \(\tilde{g}\) and \(\tilde{n}\), as direct
consequences of eqs. (\ref{rescaled-gap-eq},\ref{rescaled-n-eq}) and
(\ref{rescaled-tc-eq},\ref{density-eq-rescaled}), respectively.

\newpage\onecolumn

\end{document}